\documentclass[final,3p,times,twocolumn,fleqn]{elsarticle}

\usepackage{graphicx}
\usepackage{amssymb}
\usepackage{subfigure}
\newcommand\etal{{\it et al.}}

\journal{Physics Letters B}

\begin{document}

\begin{frontmatter}

\title{Three dimensional structure of low-density nuclear matter}

 \author[tsukuba,jaea]{Minoru OKAMOTO\corref{cor1}}
 \ead{okamoto@nucl.ph.tsukuba.ac.jp}
 \author[jaea,tsukuba]{Toshiki MARUYAMA}
 \ead{maruyama.toshiki@jaea.go.jp}
 \author[tsukuba,css]{Kazuhiro YABANA}
 \ead{yabana@nucl.ph.tsukuba.ac.jp}
 \author[kyoto]{Toshitaka TATSUMI}
 \ead{tatsumi@ruby.scphys.kyoto-u.ac.jp}

 \address[tsukuba]{Graduate School of Pure and Applied Science, University of Tsukuba, Tennoudai 1-1-1, Tsukuba, Ibaraki 305-8571, Japan}
 \address[jaea]{Advanced Science Research Center, Japan Atomic Energy Agency, Shirakata Shirane 2-4, Tokai, Ibaraki 319-1195, Japan}
 \address[css]{Center of Computational Sciences, University of Tsukuba, Tennoudai 1-1-1, Tsukuba, Ibaraki 305-8571, Japan}
 \address[kyoto]{Department of Physics, Kyoto University, Kyoto 606-8502, Japan}

 \cortext[cor1]{Corresponding author}

\begin{abstract}
We numerically explore the pasta structures and properties of low-density nuclear matter without any assumption on the geometry.
We observe conventional pasta structures, while a mixture of the pasta structures appears as a metastable state at some transient densities.
We also discuss the lattice structure of droplets.
\end{abstract}

\end{frontmatter}

\section{Introduction}
In low-density nuclear matter, which is realized in the supernovae core or in the crust of neutron stars, inhomogeneous structures of nuclear matter are expected \cite{haensel}.
With increase of density, which ranges from well below to the normal nuclear density, the shape of constituent nuclei is expected to change from spherical droplet to cylindrical rod, slab, cylindrical tube, spherical bubble, and to uniform.
These shapes are figuratively called as ``nuclear pasta''.

The density-dependence of the species and the sizes of pasta are determined by minimizing the total energy density, i.e.\ the sum of the bulk, the surface, and the Coulomb energy densities.
We can roughly assume that nuclear matter at sub-saturation density consists of a dilute gas phase and a dense liquid phase in chemical equilibrium, which determines particle densities in both phases.
Thus the bulk energy density is independent of the structure, once the volume fraction is given, but the shape and the size of the structure are determined by the balance between the surface and the Coulomb energy densities.

Among the early studies of nuclear pasta, geometrical symmetry of the structure was very often assumed, i.e.\ the Wigner-Seitz (WS) cell approximation was employed:
The whole space is divided into equivalent cells with charge neutrality, and a geometrical symmetry is assumed with a given dimensionality.
Then the shapes of the cell becomes sphere in three dimension (3D), cylinder in 2D, and plate in 1D cases.
Reflective boundary condition is imposed to the density distribution of each particle.
There is no interaction between cells due to the charge neutrality.
So all the physical quantities of matter are represented by those of a single WS cell.
Furthermore, the calculation is only in one-dimension due to the symmetry, which drastically reduces the computation.

Within this approximation and assuming uniform density distributions in each phase, analytic expressions of the surface and the Coulomb energies can be derived for a given dimensionality of the symmetry.
In 1983 Ravenhall \etal\ have shown \cite{Ravenhall83} that the configuration to give the lowest energy changes from spherical nuclear droplet (3N) to cylindrical nuclear rod (2N), nuclear slab (1NB), cylindrical bubble (2B), spherical bubble (3B) and to uniform (U) with increase of the density.
In 1984 Hashimoto \etal\ have included the Coulomb interaction among cells \cite{Hashimoto84} and have got essentially the same results.
In this work the Coulomb energy was evaluated without the WS approximation, but only simple shapes of nuclei were assumed.

On the other hand, Williams and Koonin have calculated nuclear pasta using the Thomas-Fermi model for a system in a cubic cell with periodic boundary conditions \cite{Williams}.
In this calculation no assumption was made for the structure of matter and they got essentially the same results as the studies with the WS cell approximation.
There is also a recent calculation with the Hartree-Fock theory in a cubic cell with periodic boundary conditions, giving almost the same results again \cite{Newton}.
Though the above calculations did not assume any particular structure, the size of the cell was rather small so as to include only one period of the pasta structures.
The usage of small cell brings about an implicit but strong restriction to the matter structure, suppressing the appearance of complex structures.

In this article we perform three-dimensional calculations in periodic cubic cells with sufficiently large sizes.
We will discuss how the pasta structures appear in the ground state of matter, showing their crystalline structures.
We also show some metastable states of matter which may be realized at finite temperatures.

\section{Method}
To describe the baryon interaction, we employ the relativistic mean-field (RMF) model with the Thomas-Fermi approximation \cite{maruyama}.
The RMF model deals with fields of mesons and baryons introduced in a Lorentz-invariant way.
It is rather simple for numerical calculations, but realistic enough to reproduce the bulk properties of nuclear matter.
From the variational principle we get the coupled equations of motion for the mean-fields and the Coulomb potential as
\begin{eqnarray}
-{\nabla}^{2}{\sigma}({\bf r})+m^{2}_{\sigma}{\sigma}({\bf r}) &=& g_{{\sigma}N}\left(\rho_p^s({\bf r})+\rho_n^s({\bf r})\right)\nonumber\\
 &&\hspace{-13mm}-bm_Ng_{\sigma N}^3\sigma({\bf r})^2+cg_{\sigma N}^4\sigma({\bf r})^3\label{sigma_rmf}\\
 -{\nabla}^{2}{\omega}_{0}({\bf r})+m_{\omega}^{2}\omega_{0}({\bf r}) &=& g_{{\omega}N}(\rho_{p}({\bf r})+\rho_{n}({\bf r}))\label{omega_rmf}\\
 -{\nabla}^{2}R_{0}({\bf r})+m_{\rho}^{2}R_{0}({\bf r})&=& g_{{\rho}N}({\rho}_{p}({\bf r})-{\rho}_{n}({\bf r}))\label{rho_rmf}\\
 {\nabla}^{2}V_{\rm Coul}({\bf r}) &=& e^2\left({\rho}_p({\bf r}) + {\rho}_e({\bf r})\right) \label{Coulomb_eq}
\end{eqnarray}
where $\rho_i^s({\bf r})=\langle \bar{\psi}_i({\bf r}){\psi}_i({\bf r}){\rangle}, i=p,n$ is the nucleon scalar density.
 We use the same parameters 
 as in Ref.\ \cite{maruyama} so as to compare the EOS and structural changes of pasta with or without WS cell approximation.

Equations of motion for fermions simply yield the standard relations between the densities and chemical potentials,
\begin{eqnarray}
\mu_n  &=& \sqrt{k_{{\rm F},n}({\bf r})^2+{m_N^*({\bf r})}^2} \nonumber \\
       && \hspace{15mm}+g_{\omega N}\omega_0({\bf r})-g_{\rho N}R_0({\bf r}) \label{eq:cpotB}\\
\mu_p
      &=& \sqrt{k_{{\rm F},p}({\bf r})^2+{m_N^*({\bf r})}^2} \nonumber\\
       && +g_{\omega N}\omega_0({\bf r})+ g_{\rho N}R_0({\bf r})-{V_{\rm Coul}({\bf r})} \label{eq:cpotBp} \\
{\rho_e({\bf r})}&=&-(\mu_e-{V_{\rm Coul}({\bf r})})^3/3\pi^2 \label{eq:rhoe}
\end{eqnarray}
where $m_N^*({\bf r})=m_N-g_{\sigma N}\sigma({\bf r})$ is an effective mass.
We assume here the global charge neutrality.
In case of cold catalyzed matter, the relation $\mu_n=\mu_p+\mu_e$ is further imposed.

To numerically simulate infinite matter, we use a cubic cell with periodic boundary conditions.
We divide the cell into three-dimensional grid points.
The best cell size should be as large as to include several periods of the pasta structures, and the best grid width should be as small as to attain smooth meson and fermion fields.
Due to the limitations in the memory and the CPU time, we use the cell size of $\sim 60$ fm and the grid width $|d{\bf r}| \sim 0.8$ fm.
If only one or two periods of structure appear in the calculation cell, its shape may be affected by the boundary condition and the appearance of some structures is implicitly suppressed.

Giving average densities of baryons ($\rho_B$) and electrons, initial density distributions are randomly prepared.
Then proper density distributions and meson fields are searched for.
To obtain the density distributions of baryons and electrons we introduce the local chemical potentials $\mu_a({\bf r})$ ($a=p,n,e$).
The equilibrium state is determined so that the chemical potentials are independent of the position.
An exception is the region with an empty particle density, where the chemical potential of that particle becomes higher.
We repeat the following procedures to attain uniformity of the chemical potentials.
A chemical potential $\mu_i({\bf r})$ of a baryon $i=p,n$ on a grid point ${\bf r}$ is compared with those on the six neighboring grids ${\bf r}'={\bf r}+d{\bf r}$, ($d{\bf r}=\pm d{\bf x},\pm d{\bf y},\pm d{\bf z}$).
If the chemical potential of the point under consideration is larger than that of another $\mu_i({\bf r})>\mu_i({\bf r}')$, some part of the density will be transferred to the other grid point.
This adjustment of the density is done simultaneously on all the grid points.
In addition to the above process, we adjust the particle densities between distant grid points chosen randomly so as to avoid making separate droplets with different $\mu_i$.
 
The meson fields and the Coulomb potential are obtained by solving Eqs.\ ({\ref{sigma_rmf})-(\ref{Coulomb_eq}) using the baryon density distributions $\rho_i({\bf r})$ ($i=p,n$) and the charge density distribution $\rho_p({\bf r})+\rho_e({\bf r})$.
These equations are solved numerically by a conjugate gradient method.
The electron density $\rho_e({\bf r})$ is directly calculated from the Coulomb potential $V_{\rm Coul}({\bf r})$ and the electron chemical potential $\mu_e$.
The global charge neutrality is then attained by adjusting $\mu_e$.  Above processes are repeated many times until we get convergence.
%
%
We used PRIMERGY BX900 of JAEA massively parallel computing system.
To finish a calculation of a typical case, about 10 CPU days is needed.

\section{Result}

We present here our first results with fixed proton fraction $Y_p$ for $Y_p=0.1,0.3$ and $0.5$, leaving catalyzed matter in another paper.
Shown in Fig.\ \ref{Density_dis} are the proton density distributions in cold symmetric matter (proton fraction $Y_p=0.5$).
We can see that the typical pasta phases with rod, slab, tube, and bubble, in addition to the spherical nuclei (droplet), are reproduced by our calculation in which no assumption on the structures was used.

\begin{figure}[htbp]
 \begin{center}
 \includegraphics[clip,width=0.45\textwidth]{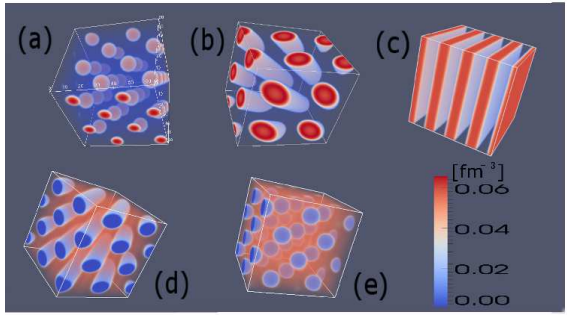}
  \caption{(color online) Proton density distributions of the ground states of symmetric matter ($Y_p=0.5$).
  Typical pasta phases are observed: (a) Spherical droplets with a fcc crystalline structure at baryon density $\rho_B= 0.01$ fm$^{-3}$.
  (b) Cylindrical rods with a honeycomb crystalline structure at 0.024 fm$^{-3}$.
  (c) Slabs at 0.05 fm$^{-3}$.
  (d) Cylindrical tubes with a honeycomb crystalline structure at 0.08 fm$^{-3}$.
  (e) Spherical bubbles with a fcc crystalline structure at 0.09 fm$^{-3}$.  \label{Density_dis}}
\end{center}
\end{figure}

\begin{figure}[htbp]
 \begin{center}
  \includegraphics[angle=-90,width=0.9\linewidth]{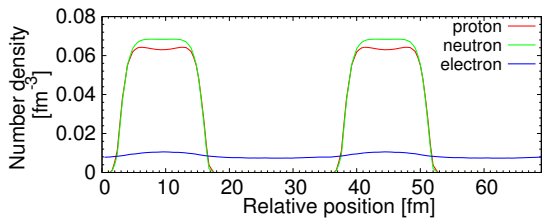}
 \end{center}
 \caption{Density profiles of proton (red), neutron (green) and electron (blue) for $Y_p=0.5$.}
 \label{dp_Yp=05}
\end{figure}

Any intermediate structures does not appear as a ground state at any density.
In a droplet, we have seen that the proton density is highest near the surface due to the Coulomb repulsion, while the neutron density distribution in the droplet is flat.
Note that baryon density outside the droplets is zero for $Y_p=0.3$ and 0.5.
The electron density 
is finite over all space but slightly localized around the droplets.
We can see this behavior of fermions for $Y_p=0.5$ and $\rho_B=0.016 \rm{fm^{-3}}$ in Fig.\ \ref{dp_Yp=05}, where plotted are the densities of proton, neutron and electron along a line which crosses through the droplets.

We show the density dependence of the energy, the total pressure, and the baryon partial pressure in Fig.\ \ref{EOS_Yp=05}.
The density dependence of these pressures is qualitatively the same as the one with the WS cell approximation.
The difference between our results and those with the WS cell approximation, in which the same RMF framework is used, is the density region of each pasta structure; density region of the rod is wider and the tube narrower in our calculation.
Figure \ref{size} shows the radius of droplets $R_d$, the lattice constant $R_{\rm cell}$, and the volume fraction $u$ of droplets.
Here, $R_d$ and $R_{\rm cell}$ are defined as follows,
\begin{eqnarray}
\frac{V}{N_d} &=& \frac{4{\pi}}{3}R_{\rm cell}^3, \\
 R_d &=& R_{\rm cell}\left(\frac{\langle{\rho}_p\rangle^2}
{\langle{\rho}^2_p\rangle}\right)^{1/3},
\label{R}
\end{eqnarray}
and $u=(R_d/R_{\rm cell})^3$, where $V$ denotes the cell volume, $N_d$ the number of droplets in the cell, and the bracket $\langle ... \rangle$ the average over the cell volume.
Comparing the present results and those with the WS cell approximation, the volume fraction shows the same behavior, but the lattice constant and the radius of the droplets are different.

\begin{figure*}[htpb]
 \begin{center}
  \subfigure{\includegraphics[height=0.3\linewidth,angle=-90]{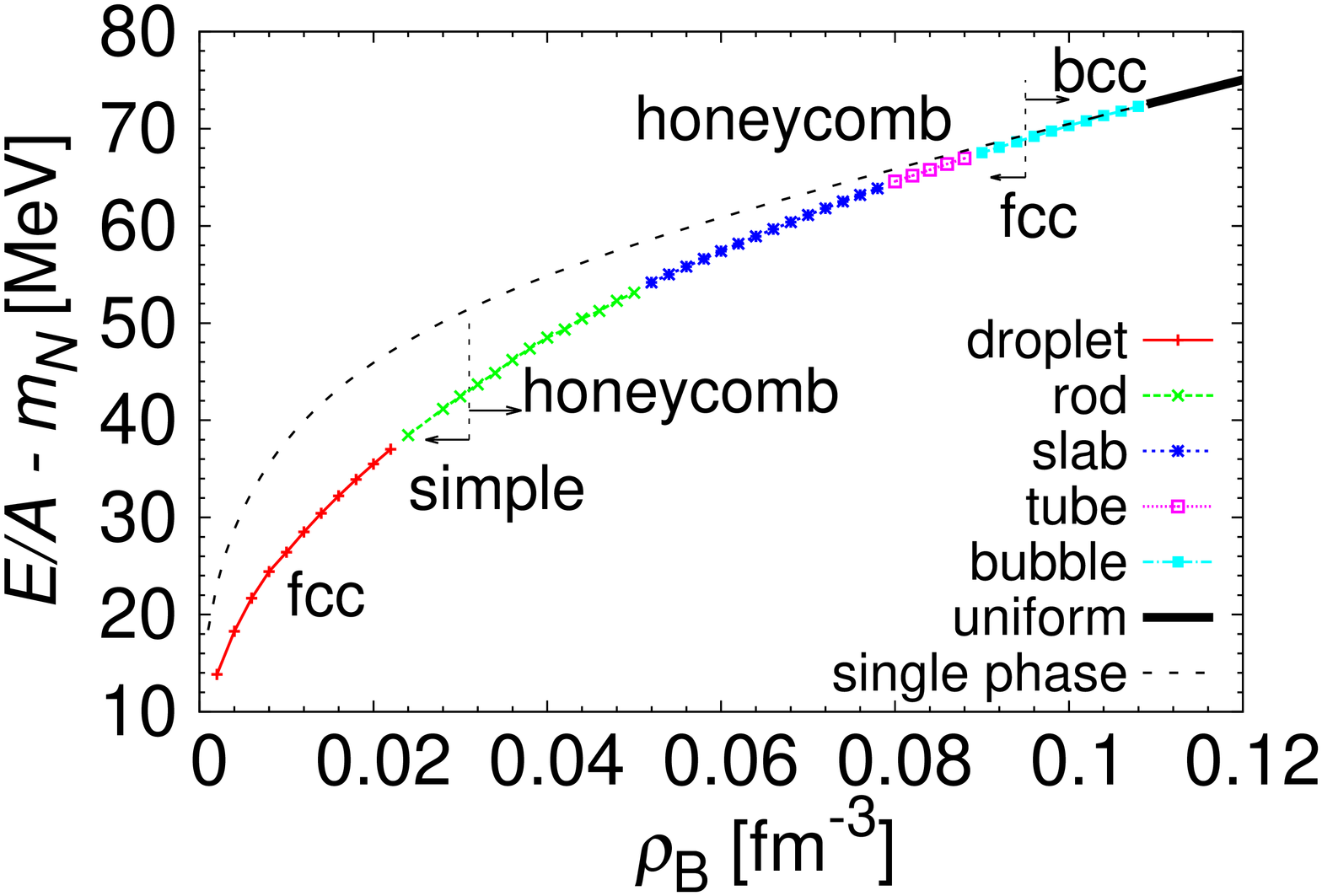}}
  \subfigure{\includegraphics[height=0.3\linewidth,angle=-90]{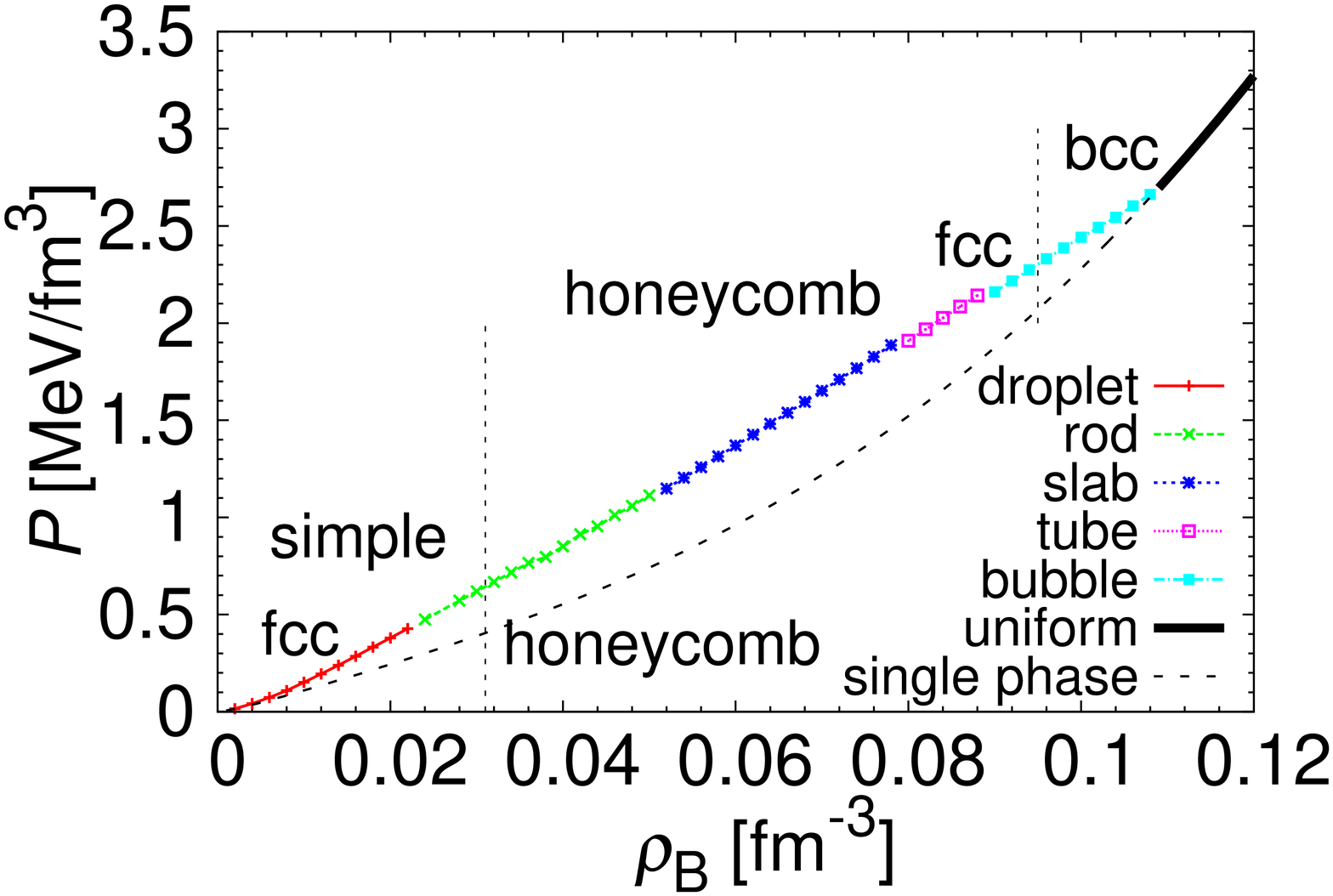}}
  \subfigure{\includegraphics[height=0.3\linewidth,angle=-90]{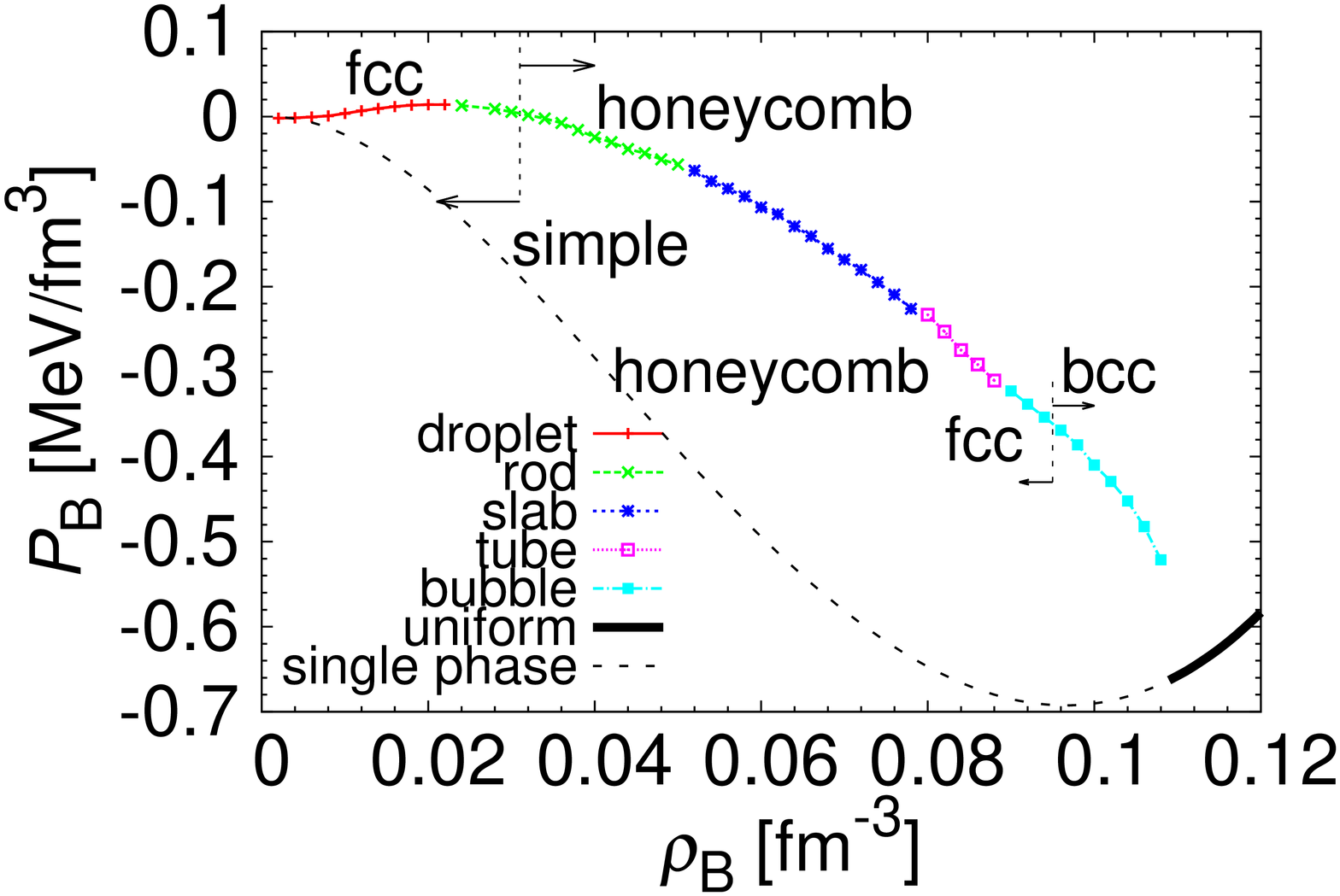}}
  \caption{(color online) From the left, energy, pressure, and baryon partial pressure of symmetric matter.
  Red lines indicate droplet, green rod, blue slab, magenta tube, cyan bubble, and solid black uniform, respectively.
  Dotted lines are the EOS in the case of single phase}
 \label{EOS_Yp=05}
 \end{center}
\end{figure*}

When a spherical nucleus becomes large, it becomes unstable against fission.
It is expressed as the Bohr-Wheeler condition \cite{Preston}, $E_{\rm Coul}^{(0)}>2E_{\rm surf}$, where $E_{\rm Coul}^{(0)}$ is the Coulomb energy of a nucleus.
In the WS cell approximation, the Coulomb energy in a cell is expressed using the Coulomb energy of a nucleus as $E_{\rm Coul}\approx E_{\rm Coul}^{(0)}\left(1-3u^{1/3}/2\right)$.
On the other hand, the condition of optimum nuclear size gives $E_{\rm surf}=2E_{\rm Coul}$.
From these equations, the appearance of non-spherical nucleus in nuclear matter due to the fission instability has been expected for the volume fraction $u>1/8$.
However, in our calculation, the structural change from droplet to rod occurs around a volume fraction $u=0.2$ (see the value at $\rho_B\approx 0.02$ fm$^{-3}$ in Fig.\ \ref{size}).
The relation between the Coulomb energies of a cell and that of a nucleus has been derived by using a uniform background electrons and uniform baryon density in a nucleus.
The effect of the screening by charged particles, which is naturally included in both the WS cell calculation in Ref.\ \cite{maruyama} and our present calculation, may be one of the origins of this difference.
Also the difference of the droplet surface may be the origin, since they used the compressible liquid drop model with a sharp surface, while our droplets have a diffuse surface.


There are some differences between our calculation and the one with the WS cell approximation.
Calculating in a large cell which has several periods of pasta structures, the interactions among the unit structures are properly included.
By comparing both results, we can see almost the same behavior of the volume fraction, but slightly larger in our case.

\begin{figure}[htbp]
 \begin{center}
  \includegraphics[width=0.7\linewidth,angle=-90]{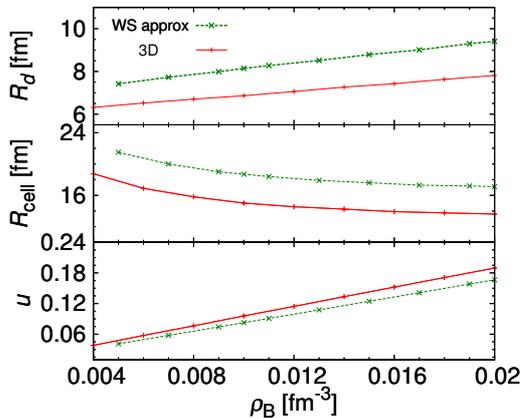}
 \end{center}
 \caption{Density dependence of the radius, the lattice constant and the volume fraction.
  Red lines are our results and green lines are those with the WS cell approximation}
 \label{size}
\end{figure}

One point unlike the conventional results emerged in the crystalline structure of droplets.
In our calculation, it emerges as a face-centered cubic (fcc) lattice, while it has been regarded to take a body-centered cubic (bcc) lattice in the previous studies \cite{oyamatsu,nakazato}.
Crystalline structures in the bcc and fcc lattices give rise to a subtle difference of the Coulomb energy, which amounts to about $0.2$--$0.8$ MeV.
The ratio of the Coulomb energy for total energy difference is about $20\%$.
In other words, most of energy differences come from the bulk energy of the nuclear matter.

\begin{table}[htbp]
 \begin{tabular}{cccccc}
  \hline
  $\rho_B$ [$\rm fm^{-3}$] &   0.012  &  0.014  &  0.016  &  0.018  &  0.020 \\
  \hline
  ${R_d(\rm fcc) [fm]}$     &  6.86   &  7.04  &  7.23  &  7.61  &  7.79 \\
  ${R_d(\rm bcc) [fm]}$    &   6.99  &  7.18  &  7.36  &  7.75  &  7.92 \\
  \hline
 \end{tabular}
 \caption{Density dependence of droplet radii for the fcc and bcc
 lattices.}
 \label{dif_f_b}
\end{table}

Table \ref{dif_f_b} shows the baryon-density dependence of radii of droplets in the cases of the fcc and bcc lattices.
The radii of droplets are different even if their baryon densities are the same.
In Refs.\ \cite{oyamatsu,nakazato}, it is reported that the bcc crystalline structure of droplets is realized in the ground state at low densities, while the fcc crystalline structure appears in our calculation.
This difference may partially come from that they compare the bcc and fcc crystalline structures using droplets of the same size, while in our case they have different droplet sizes.
In the QMD calculations \cite{QMD1,QMD2} that precede the present calculations without assuming structure not only for static but also for dynamical aspects of nuclear matter, droplets form the bcc crystalline structure.
  This difference might come from the treatment of electrons:
  In the QMD calculation, uniform electron distribution has been assumed, while in our calculation, inhomogeneous electron distribution is attained consistently.
  We have performed another calculation with uniformly distributed electrons.
  However, the result did not change and the fcc structure is more favored.
  Thus it is confirmed that the charge screening by electrons does not very much affect the crystalline structure \cite{maruyama}.
  This difference of the crystalline structure between the QMD calculation and the present calculation remains as a future problem.

\begin{figure*}[htbp]
 \begin{center}
  \subfigure{\includegraphics[height=0.28\linewidth,angle=-90]{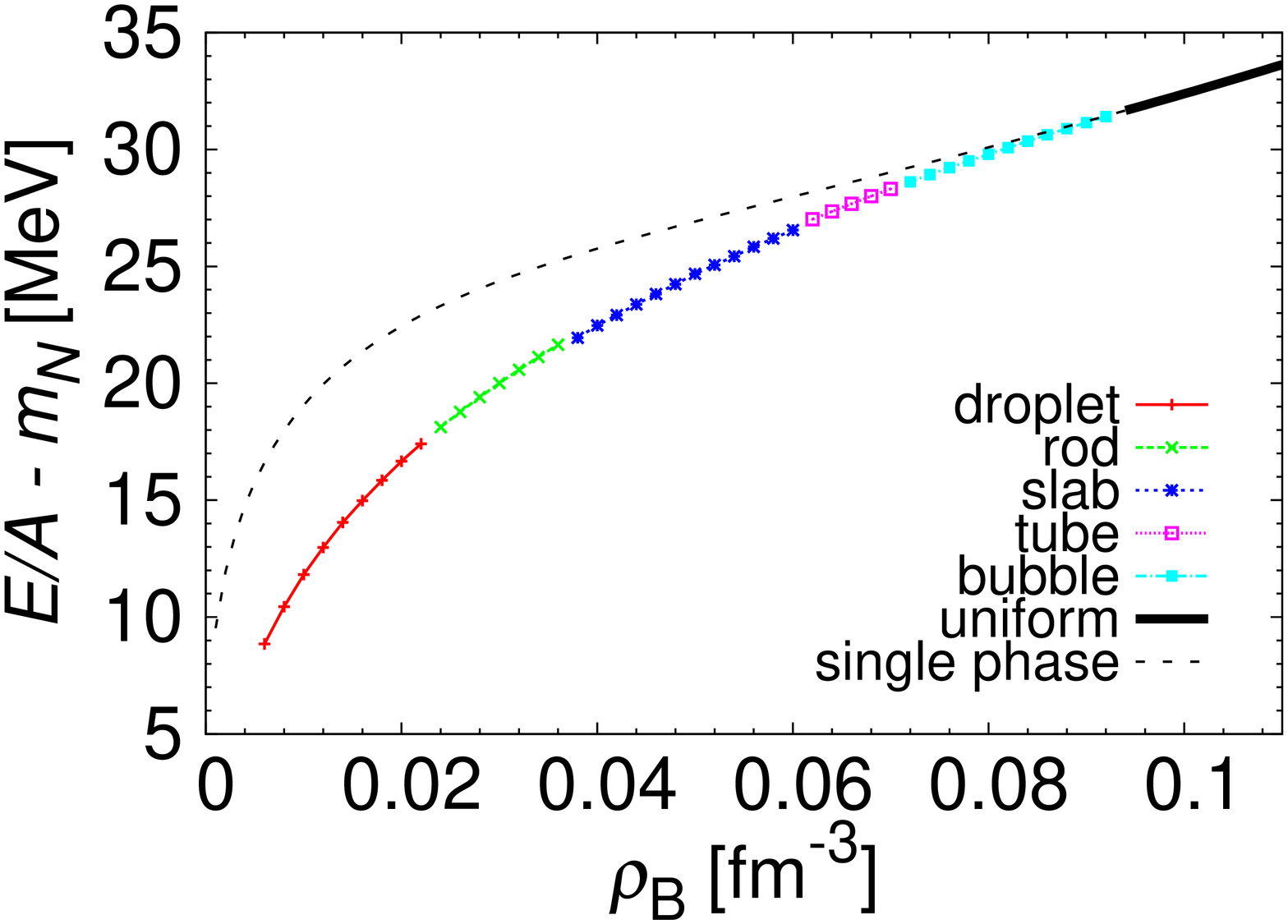}}
  \subfigure{\includegraphics[height=0.28\linewidth,angle=-90]{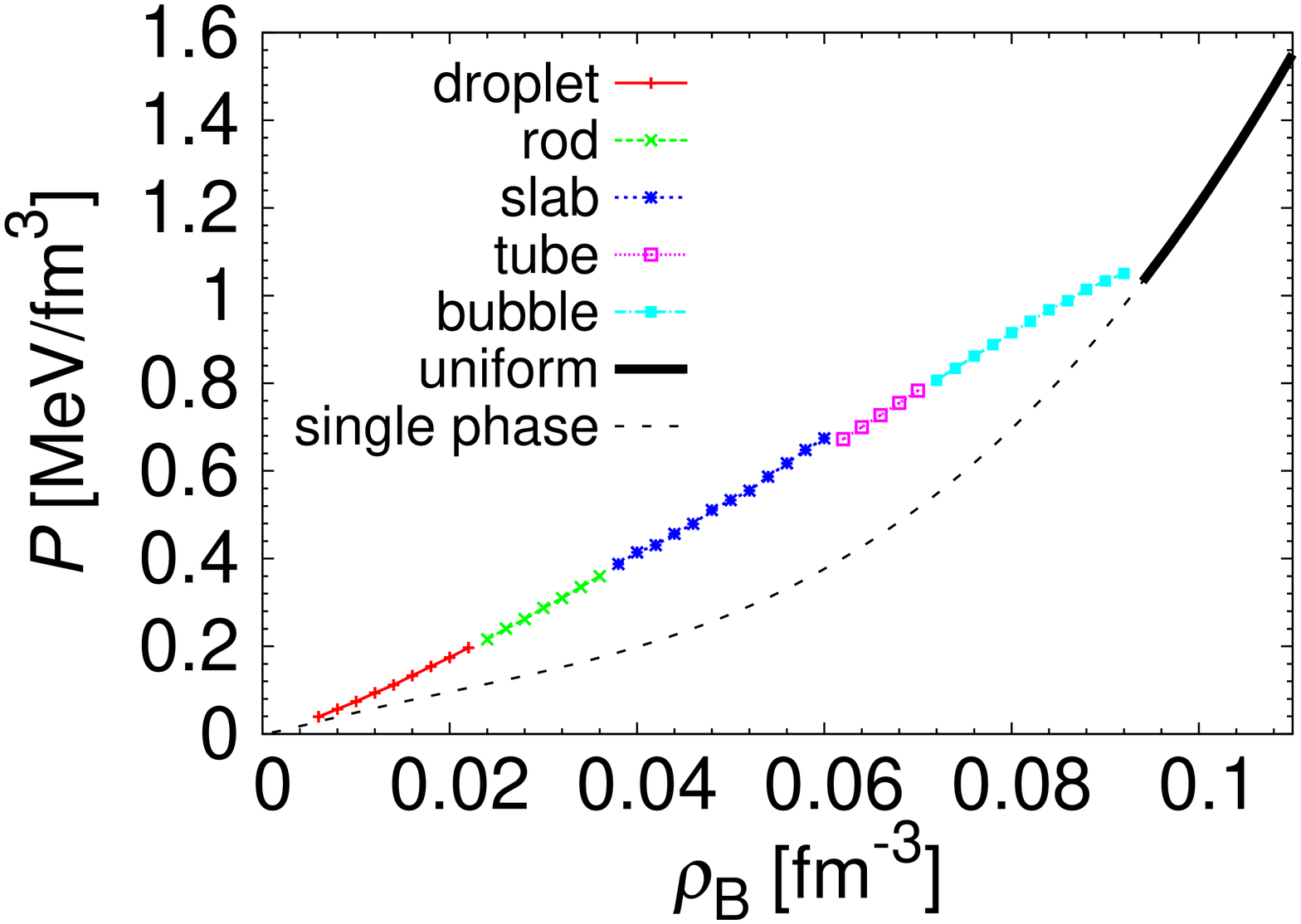}}
  \subfigure{\includegraphics[height=0.28\linewidth,angle=-90]{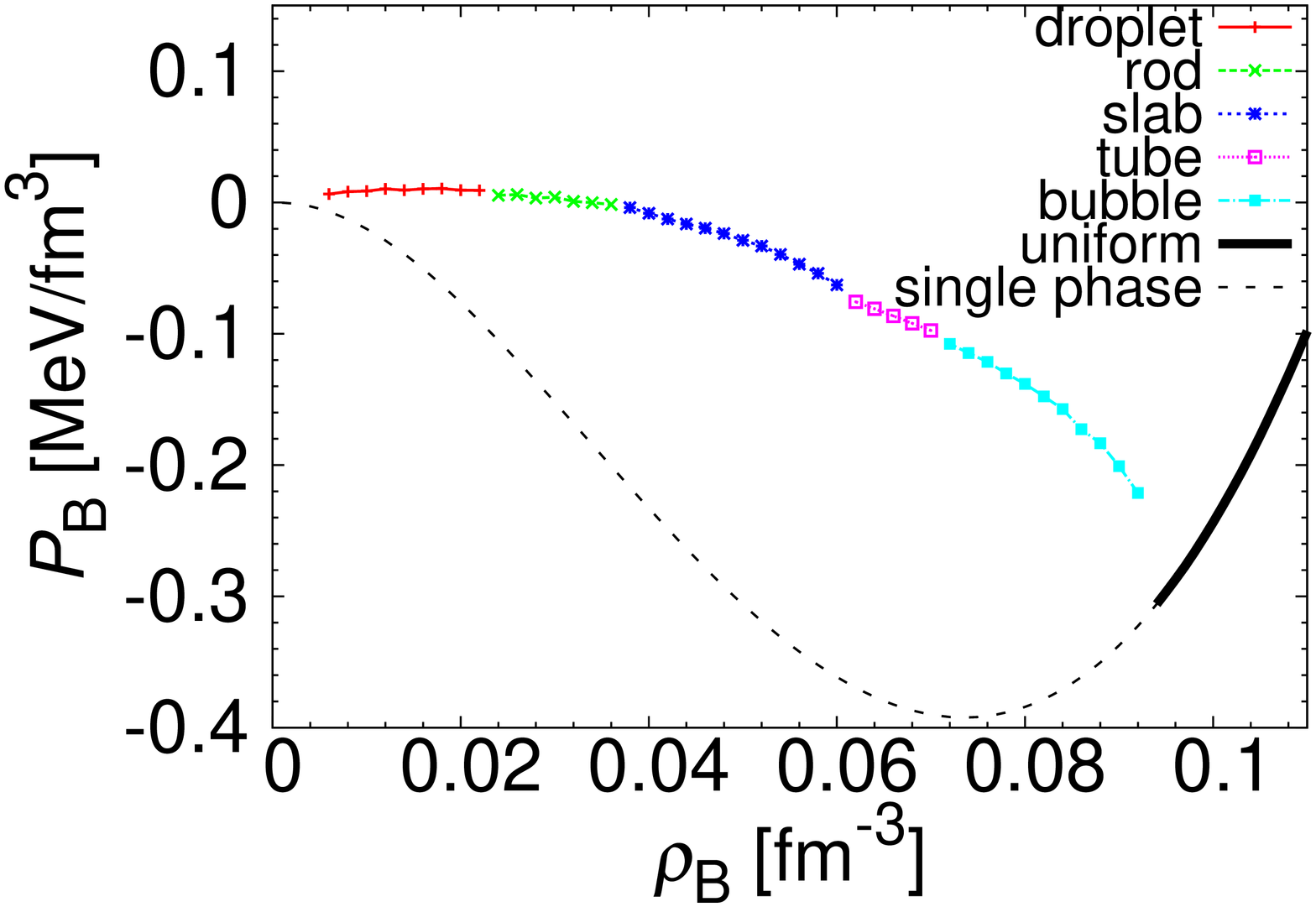}}
  \caption{(color online) 
  Same as Fig.\ \ref{EOS_Yp=05} for asymmetric matter with $Y_p=0.3$.}
 \label{EOS_Yp=03}
 \end{center}
\end{figure*}
\begin{figure*}[htbp]
 \begin{center}
  \subfigure{\includegraphics[height=0.28\linewidth,angle=-90]{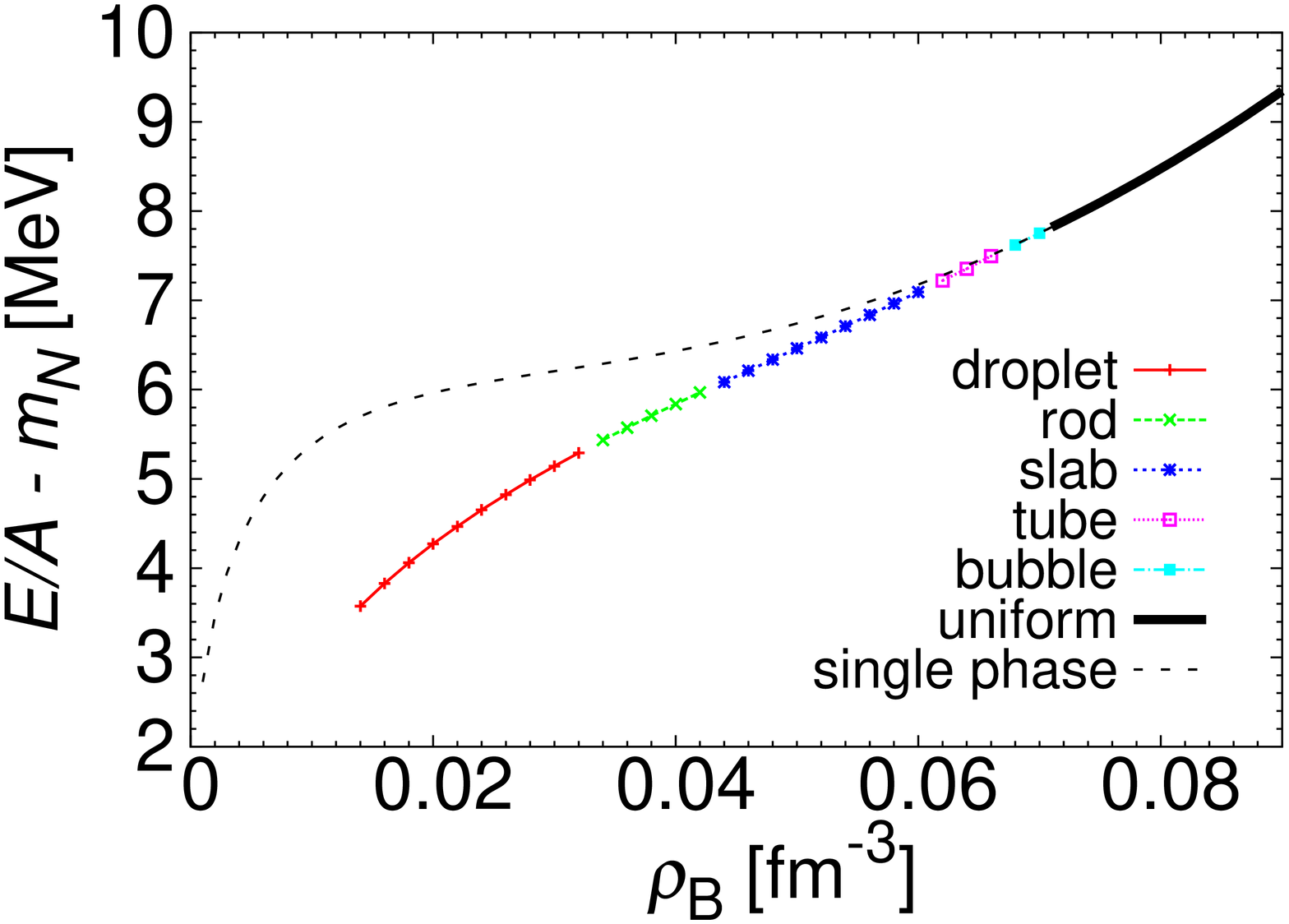}}
  \subfigure{\includegraphics[height=0.28\linewidth,angle=-90]{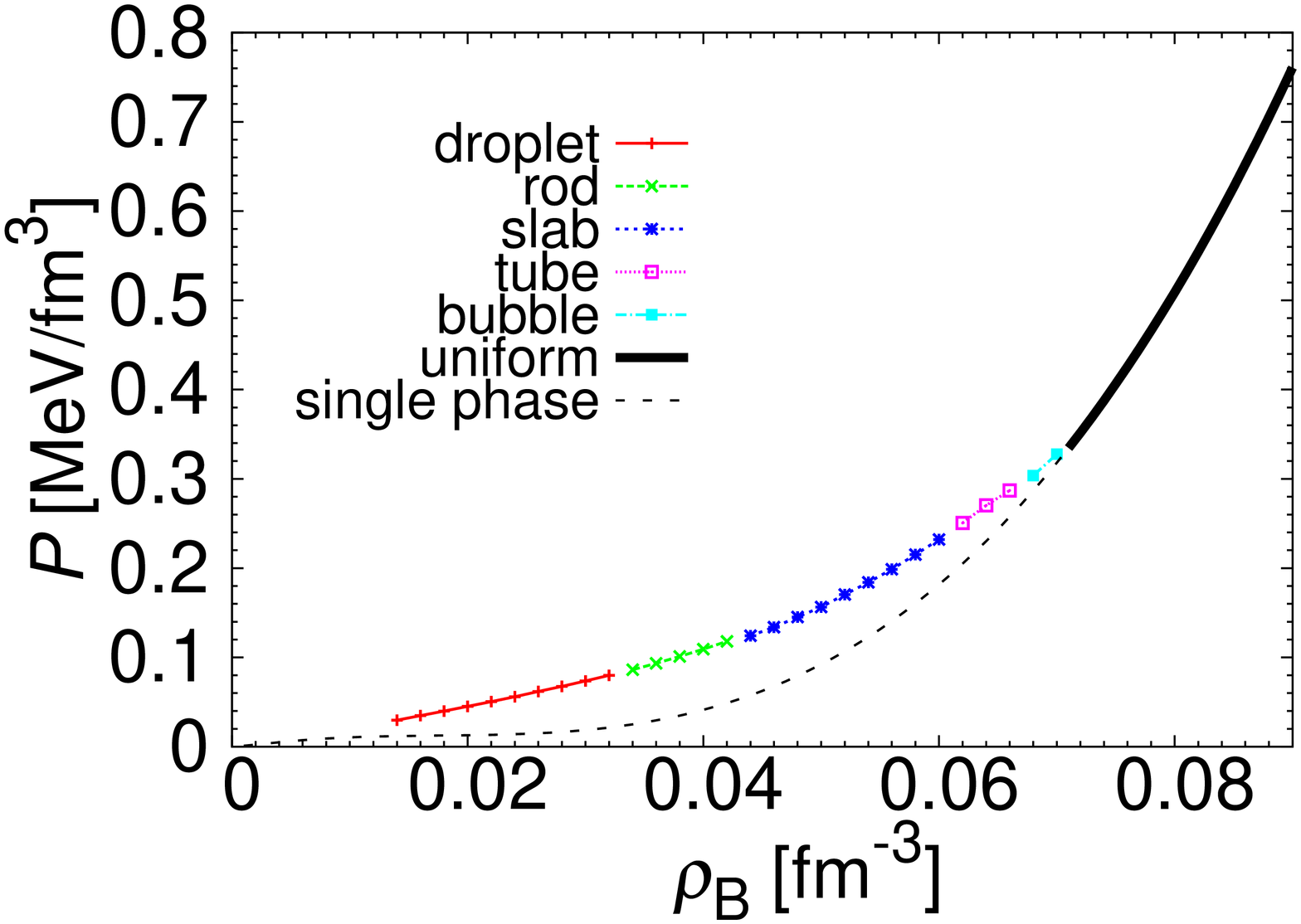}}
  \subfigure{\includegraphics[height=0.28\linewidth,angle=-90]{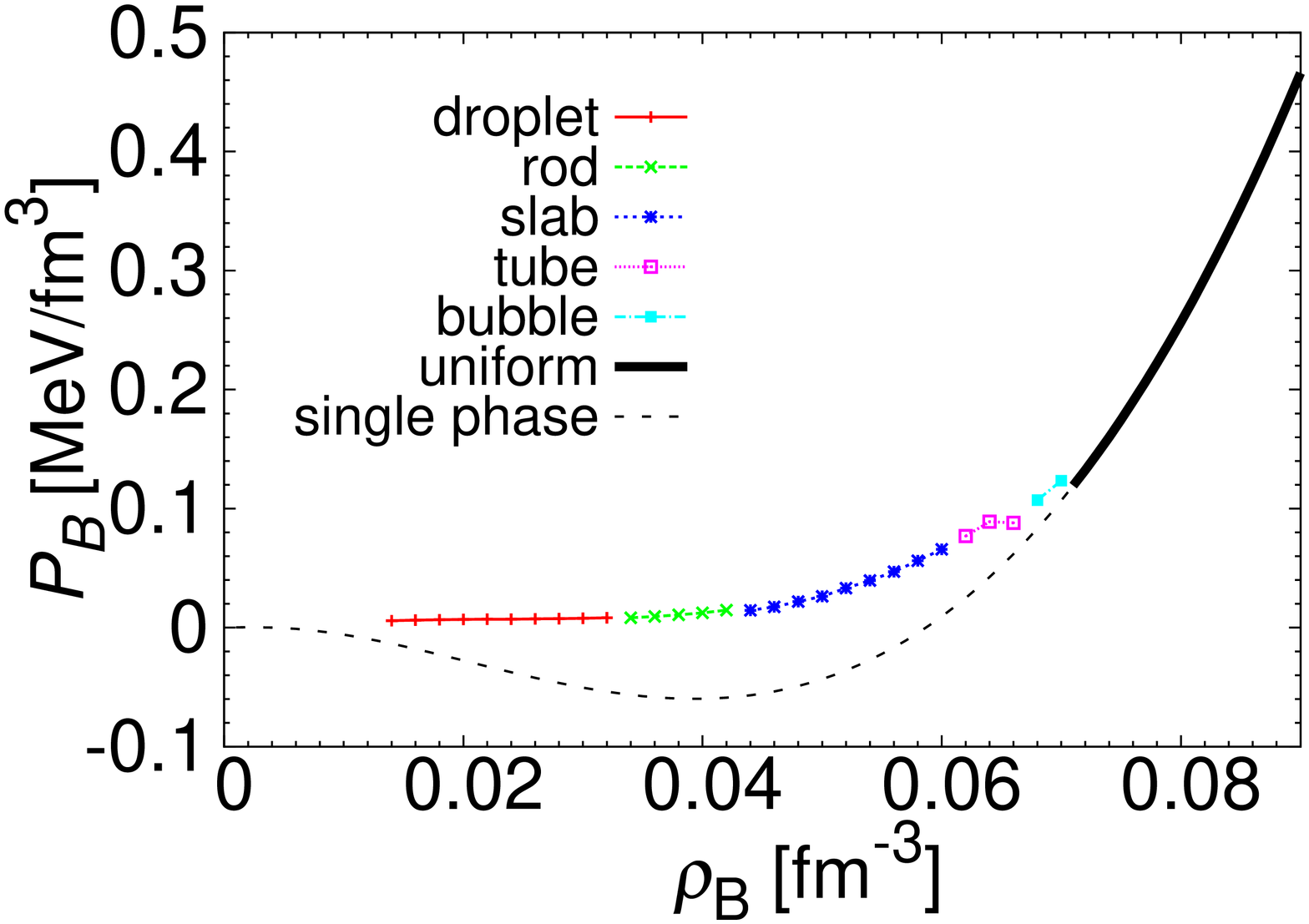}}
  \caption{(color online) 
  Same as Fig.\ \ref{EOS_Yp=05} for asymmetric matter with $Y_p=0.1$.}
 \label{EOS_Yp=01}
 \end{center}
\end{figure*}

In Figs.\ \ref{EOS_Yp=03} and \ref{EOS_Yp=01}, we show the density dependence of the energy, the total pressure and the baryon partial pressure for $Y_p= 0.3$ and 0.1, which are roughly relevant to the supernova matter and the neutron star crust.
We obtain the typical pasta structure as ground states for any proton fraction above 0.1.
Also in the cases of $Y_p$=0.3 and 0.1, the fcc lattice of droplets is energetically more favorable than the bcc.
For rod and tube, the simple lattice is more favorable than the honeycomb lattice.
In Fig.\ \ref{dp_Yp=01} we plot the density profile of proton and neutron for $Y_p=0.1$ with baryon density $0.016 \rm{fm^{-3}}$ along a line which crosses through the droplets.
The neutron density is finite at any point: the space is filled by dripped neutrons, which is in contrast to the case with higher $Y_p$ where the neutron density is zero outside the nucleus.
\begin{figure}[htbp]
 \begin{center}
  \includegraphics[angle=-90,width=0.9\linewidth]{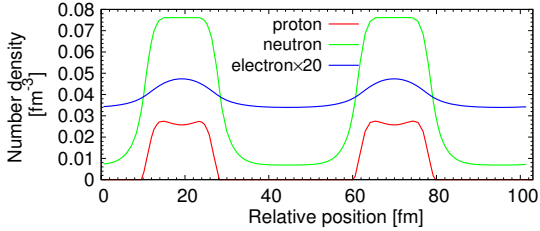}
 \end{center}
 \caption{Density profiles of proton (red) and neutron (green) and
 electron (blue) for $Y_p=0.1$. To show clearly the distribution of
 electron, the electron density is magnified 20 times.}
 \label{dp_Yp=01}
\end{figure}

\begin{figure}[htbp]
 \begin{center}
 \includegraphics[width=0.9\linewidth]{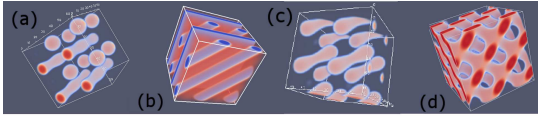}
  \caption{Proton density distributions with complex structures ($Y_p=0.5$).
  (a) mixture of droplet and rod, 0.022 fm$^{-3}$,
  (b) slab and tube, 0.068 fm$^{-3}$ ;
  (c) dumbbell like structure, 0.018 fm$^{-3}$ ;
  (d) diamond like structure, 0.048 fm$^{-3}$.
 \label{meta}}
\end{center}
\end{figure}
On the way of searching for the ground-state structures, we sometimes observe exotic structures which are energetically metastable.
Figure \ref{meta} (a) shows a mixed structure of droplet and rod at $\rho_B=0.022$ fm$^{-3}$.
Similarly, (b) is a mixed structure of slab and tube at 0.068 fm$^{-3}$.
These structures appear around densities where the structures change.
If these structures had appeared as ground states, transition from droplet to rod, and from slab to tube would happen more smoothly by way of those intermediate structures.
In fact, the QMD calculations have reported some intermediate structures of droplet and rod, slab and tube as ground states \cite{QMD1}.

We have observed more exotic structures:
Panel (c) at 0.018 fm$^{-3}$ looks like dumbbells which have been reported to appear in the dynamical compression of matter \cite{QMD_supernova}.
Finally, panel (d) at 0.044 fm$^{-3}$ is a diamond structure, which resembles double-diamond structure studied with a compressible liquid drop model \cite{nakazato,matsuzaki}.
These complex structures are difficult to be the ground states since they have larger surface area than simple pasta structures.
At finite temperatures, however, these structures might contribute to the Boltzmann ensemble.

\section{Summary}

We have numerically explored inhomogeneous structures and properties of low-density nuclear matter using the RMF model and the Thomas-Fermi approximation.
Without any assumption on the geometry, we have carried out fully three-dimensional calculations on large cubic cells with periodic boundary conditions.
With increase of density, which ranges from well below to the normal nuclear density, we have observed that the ground state of matter shows the typical pasta structures.
More complex structures like ``diamond'' structure, ``dumbbell'' structure, and mixtures of two types of pasta structures appear as metastable states at some transient densities.
As for the crystalline structures for the droplet and the bubble structures, fcc lattice has been more suitable than bcc lattice, which is different from the previous studies.
For neutron star crust and supernovae matter, we should explore these structures and properties for $\beta$-equilibrium and extend to finite temperature.
At finite temperatures, complex structures might contribute to the Boltzmann ensemble.
We can apply these contributions to the mechanism of glitch and cooling process of neutron stars and thermal and mechanical properties of supernovae.
To explore $\beta$-equilibrium matter, we need more calculation space.
But there are some problems in calculation performance.
As mentioned before, to finish a calculation of a typical case, many CPU time is needed.
Most of the CPU time is consumed by the part to get the uniformity of chemical potentials and its iteration.
To reduce the CPU time, we must improve the method and calculation code of this part.
More realistic calculation can be done by, for example, including gradient terms of the density to improve the description of the surface properties \cite{gradient}.

This work is supported by the project {\it ``Research on the Emergence of Hierarchical Structure of Matter by Bridging Priticle, Nuclear and Astrophysics in Computational Science,'' 2008--2013}, of the Grant-In-Aid for Scientific Research on Innovative Areas, MEXT Japan.

\end{document}